# Optimal Addition of Images for Detection and Photometry

Philippe Fischer

AT&T Bell Laboratories, 600 Mountain Ave., 1D-316, Murray Hill, NJ 07974
Email: philf@physics.att.com

and

Greg P. Kochanski

AT&T Bell Laboratories, 600 Mountain Ave., Murray Hill, NJ 07974
Email: gpk@physics.att.com

## ABSTRACT

In this paper we describe weighting techniques used for the optimal coaddition of CCD frames with differing characteristics. Optimal means maximum signal-to-noise (s/n) for stellar objects. We derive formulae for four applications: 1) object detection via matched filter, 2) object detection identical to DAOFIND, 3) aperture photometry, and 4) ALLSTAR profile-fitting photometry. We have included examples involving 21 frames for which either the sky brightness or image resolution varied by a factor of three. The gains in s/n were modest for most of the examples, except for DAOFIND detection with varying image resolution which exhibited a substantial s/n increase. Even though the only consideration was maximizing s/n, the image resolution was seen to improve for most of the variable resolution examples. Also discussed are empirical fits for the weighting and the availability of the program, WEIGHT, used to generate the weighting for the individual frames. Finally, we include appendices describing the effects of clipping algorithms and a scheme for star/galaxy and cosmic ray/star discrimination.

*Subject headings:* CCD, coaddition, optimal filter, signal-to-noise

## 1. Introduction

For many astronomical applications it is necessary to break up long CCD exposures into a series of shorter ones. Typical reasons for this are: insufficient dynamic range, removal of cosmic rays, and ultra deep image construction via the shift-and-stare technique (Tyson 1990), etc. Therefore, frames of a single field can extend over many observing runs, during which time the conditions (i.e. sky brightness, seeing, sky transparency) and detector parameters (i.e. scale, gain, readout noise and quantum efficiency) can change. These considerations have motivated us to devise a program (see §5. for availability) designed to maximize the signal-to-noise ratio (s/n) of a stellar image on a coadded CCD frame. We will demonstrate how to weight the individual frames such that detection and measurement of faint objects is optimally enhanced. Our techniques are strictly linear so that the images remain suitable for quantitative measurement and modeling.

In §2. we discuss preliminary measures which must be taken before weight calculation and image combination are possible. In §3. we describe a technique to maximize s/n for the detection of faint sources using convolution with both a matched filter of infinite extent (§3.1.), and a lowered truncated Gaussian (§3.3.). In §4. we describe image combination for both aperture (§4.1.) and profile-fitting photometry (§4.3.). We have also included two appendices, one describing a strategy for clipping algorithms and the other outlining a techniques for star/galaxy and star/cosmic ray discrimination.

## 2. Image Preparation

In the following derivations, it is assumed that each individual CCD image has the same scale (i.e. a pixel spans the same angular extent on the sky). If images were taken with different CCDs and/or telescopes then this will not, in general, be true. It is, however, a simple matter to rescale the individual images to a common value using, for example, the IRAF[1] (Tody 1986) task MAGNIFY. We also assume that the frames are registered with respect to one another (the IRAF tasks GEOMAP and GEOTRAN can be used for this).

As an aside, binning of pixels (i.e. demagnification) preserves the white stationary random character that we will require (i.e. where the noise spatial autocorrelation function

---

is a delta function with constant size across the image) of the photon statistics. Magnifying an image, or interpolation, however, will result in short-range (high spatial frequency) correlations between pixels.

We further assume that the frames will be *added*, or, equivalently averaged. Appendix A describes the effects of employing a clipping algorithm to improve cosmic ray rejection. One argument in favor of median or clipping algorithms is that real CCD images have non-Gaussian noise-like bad pixels, cosmic rays, bad columns, and uncorrected flat field gain irregularities.

## 3. Detection

### 3.1. Matched Filter

An astronomical CCD image is generally a combination of signal (stars, galaxies etc.) and sky (scattered light in the atmosphere, unresolved sources) with noise (photon shot, readout, and flat-field). The optimal means of detecting a signal of known shape in stationary white noise is to convolve the image with the true signal [i.e. the stellar point spread function (PSF), or perhaps a galaxy profile]. This technique is known as a matched filter (Whalen 1971, p. 173). One then, typically, searches the convolved image for local maxima. If the value of a given local maximum exceeds a threshold, it is considered to be a positive detection. The threshold value is calculated to reduce the probability of false detections to an acceptable level while not rejecting too many real objects. This is similar to the approach used by the FOCAS software package (Jarvis & Tyson 1981, Valdes 1982), which uses, as a default, a hardwired star-like PSF. FOCAS will also implement an exact matched filter (although truncated at finite radius) by inputting the stellar PSF from the task AUTOPSF back into the catalog for a second pass of DETECT.

Unfortunately, the condition of stationary white noise is not exactly satisfied for astronomical observations limited by photon statistics; the variance increases with the signal. However, for sufficiently faint objects, this approximation is quite good, and this is the regime we are most interested in optimizing for detection. Our goal will be to derive a method of coadding CCD frames in such a way as to maximize the signal-to-noise ratio (s/n) of the local maxima. This will be achieved by calculating a relative weight for each individual frame prior to coaddition.

Under most observing conditions, a circular Gaussian is a good approximation to a stellar PSF, at least near the center (King 1971) (see the conclusion for a description of what happens with a more realistic PSF). With this assumption, the signal on the coadded frame, in digital unit (DU) per pixel is:

$$S(x,y) = \sum_{i=1}^{n} \frac{w_i I_i}{2\pi\sigma_i^2} e^{-[(x-x_0)^2 + (y-y_0)^2]/2\sigma_i^2}, \tag{1}$$

where $n$ is the number of frames, $w_i$ is the weight for frame $i$, $I_i$ is the total observed flux for a star in DU, including extinction and clouds, $\sigma_i$ is measured in pixels, and $x_0$ and $y_0$ are the coordinates of the star's center.

We take the variance of the signal to be:

$$[N(x,y)]^2 = \sum_{i=1}^{n} \left[ \frac{w_i^2 I_i}{2\pi\sigma_i^2 g_i} e^{-[(x-x_0)^2 + (y-y_0)^2]/2\sigma_i^2} + \frac{w_i^2 B_i}{g_i} + \frac{w_i^2 h_i^2}{g_i^2} \right], \tag{2}$$

where $B_i$ is the mean sky value in DU/pixel, $g_i$ is the gain [in $e^-$ per DU], and $h_i$ is the rms readout noise in electrons. The following assumptions have been made: 1) the different noise sources are uncorrelated, 2) the noise at a given position is uncorrelated on the different frames, 3) Poisson noise can be approximated as Gaussian, implying sufficient electrons, 4) noise resulting from imperfect flat fielding is negligible, 5) noise caused by fringing due to sky emission lines is negligible, and 6) noise resulting from interpolation errors incurred during image registration is negligible. Additional Gaussian noise terms can be easily incorporated into the readout noise term.

A CCD detects fluxes integrated over the area of a pixel; thus we cannot measure the true peak values of the local maxima. In order to obtain the flux in a pixel we convolve our signal with a square function:

$$C(x,y) = 1 \text{ if } -0.5 \leq x, y \leq 0.5 \tag{3a}$$
$$= 0 \text{ otherwise.} \tag{3b}$$

The signal portion of the flux in a pixel with center at (x,y) is

$$S_p(x,y) = C * S = \sum_{i=1}^{n} \frac{w_i I_i}{4\Psi_i(x)\Psi_i(y)}, \tag{4}$$

where we have defined

$$\Psi_i(x) = \mathrm{Erf}\left(\frac{x' + 0.5}{\sqrt{2}\,\sigma_i}\right) - \mathrm{Erf}\left(\frac{x' - 0.5}{\sqrt{2}\,\sigma_i}\right), \tag{5}$$

for convenience, $x' = x - x_0$ and $y' = y - y_0$. As it represents a pixellated image, the function $S_p$ has meaning only at integer values of $x$ and $y$. Similarly, the variance in a pixel with center at (x,y) is:

$$N_p(x,y)^2 = C^2 * N^2 = \sum_{i=1}^{n}\left(\frac{B_i w_i^2}{g_i} + \frac{w_i^2 h_i^2}{g_i^2} + \frac{w_i^2 I_i}{4 g_i \Psi_i(x)\Psi_i(y)}\right) \tag{6}$$

The matched filter approach requires convolving the coadded image with a filter shaped like $S_p$, thereby suppressing fluctuations which occur at other frequencies. The value of the maximum pixel on the filtered coadded frame is the discrete convolution of $S_p$ with itself evaluated at the pixel containing $(x_0, y_0)$:

$$S_c = (S_p \star S_p)_0 =$$

$$\sum_{x=-\infty}^{x=\infty}\sum_{y=-\infty}^{y=\infty}\sum_{i=1}^{n}\sum_{j=1}^{n}\frac{I_i I_j w_i w_j}{4}\Psi_i(x)\Psi_i(y)\Psi_j(x)\Psi_j(y). \tag{7}$$

We can remove two of the sums analytically, if we approximate the exact discrete convolution with a continuous convolution:

$$S_c \approx (S_p * S_p)_0 =$$

$$\sum_{i=1}^{n}\sum_{j=1}^{n}\frac{I_i I_j w_i w_j}{4}\left[\mathrm{Erf}\left(\frac{x_0'}{\sigma'}\right) + \mathrm{Erf}\left(\frac{1 - x_0'}{\sigma'}\right) + \mathrm{Erf}\left(\frac{y_0'}{\sigma'}\right) + \mathrm{Erf}\left(\frac{1 - y_0'}{\sigma'}\right)\right], \tag{8}$$

where: $x_0' = x_0 - \mathrm{Int}(x_0)$, $\quad y_0' = y_0 - \mathrm{Int}(y_0)$, $\quad$ and $\quad \sigma' = \sqrt{4\sigma_i^2 + 4\sigma_j^2}$. The variance of the maximum pixel on the convolved frame is the discrete convolution of $S_p^2$ with $N^2$, evaluated at the nearest pixel center to $(x_0,y_0)$, but as it involves a quintuple sum, we will again use the continuous approximation

$$N_c^2 = (S_p^2 \star N^2)_0 \approx (S_p^2 * N^2)_0 =$$

$$\sum_{i=1}^{n}\sum_{j=1}^{n}\sum_{k=1}^{n}\frac{I_i I_k w_i w_j^2 w_k}{2\pi g_j^2(\sigma_i^2 + \sigma_k^2)}\left[(h_j^2 + g_j b_j) + g_j I_j \mathrm{Erf}\left(\frac{\sqrt{\sigma_i^2 + \sigma_k^2}}{2\sqrt{2}\sqrt{\sigma_i^2\sigma_j^2 + \sigma_i^2\sigma_k^2 + \sigma_j^2\sigma_k^2}}\right)^2\right]. \tag{9}$$

The s/n of the matched filter is then $S_c/N_c$.

Our goal is to choose the $w_i$ such that the s/n is maximized. In general, the optimized weights will depend on the brightness of the objects of interest, ranging from a faint object limit (compared to the sky brightness) to a limit where the sky brightness is negligible. We have written a program which uses the Downhill Simplex Method described in Press et al. (1986) which finds the set of weights which maximize the s/n (see §5.) In most cases, one will want to optimize the weighting for detection of objects near a detection threshold.

### 3.2.  Examples

The left side of Fig. 1 shows the values of $w_i$ for four simulations of 21 frames. Each has readout noise $h_i = 5e^-$, gain $g_i = 3e^-/\mathrm{DU}$, sky brightness $B_i = 1000\,\mathrm{DU}$, and Gaussian half-width $\sigma_i$ ranging from 1.0 to 3.0 pixels in steps of 0.1. The star is assumed to be centered on a pixel. The results are summarized in Table 1; column 1 is the stellar intensity, column 2 is the maximized s/n, column 3 is the full-width-half-maximum (FWHM) of an image on the coadded frame, column 4 is the equal weight s/n and column 5 is the equal-weight FWHM. The low brightness frames behave as expected; s/n is maximized when the higher resolution frames are weighted more heavily than the lower resolution frames. The solid line in Fig. 1 represents the asymptotic weighting as the signal tends to zero, demonstrating that the weighting does not vary much for a fairly large range of stellar fluxes, from zero to well above the usual detection thresholds (i.e. typically s/n > 4). As the stellar brightness increases, the slope of the $w_i$ vs $\sigma_i$ relationship flattens. The relationship actually turns over in the highest signal case such that the maximum weighted frames are those with intermediate resolution. In the high signal regime our filter is no longer a true "matched filter" since our noise is not stationary. While this regime is little more than a curiosity, we have included in Fig. 1 a dashed line representing the weighting for zero sky and readout noise.

The two main questions of interest are: how significantly have the s/n and image resolution improved over the equal-weight case? The signal-to-noise ratios have improved by a few percent while the FWHM for the lowest s/n example has decreased by about 10%. These are both fairly modest gains despite a factor of three range in the seeing on the individual frames.

In a real CCD image, the stars on a given frame will have centers falling randomly with respect to the center of a pixel. The optimal weighting is somewhat dependent on exactly where a star's center lies, and this dependence is stronger if the sampling is reduced. To test this dependency, we have rerun the $I_i = 200\,\mathrm{DU}$ example, but this time with a maximally off-centered star (the star center is positioned on the intersection of four pixels). The maximum difference in the optimal weights for this off-centered case vs the centered case was less than 7%. More importantly, when we applied the optimal weights from the *centered* case to the non-centered case, the observed s/n improvement was similar; the s/n improved from 7.0 to 7.3 (equal weights vs. optimal weights) as compared to 7.3 to 7.7 for the centered case. The final FWHM in this case is 3.18 pixels, which is typical of many observing runs, implying that the optimal centered weights will be adequate for many real situations. In more poorly sampled images, it may be advantageous to use an average of the two extremes, centered and maximally decentered.

Another interesting case to examine is constant seeing with a changing sky level. Choosing the readout noise $h_i$ and gain $g_i$ as above, with $\sigma_i = 1.0$ pixels and the sky varying from 1000 DU to 3000 DU in steps of 100 DU, we obtain the right panel of Fig 1. The symbols are analogous to the left side, and the relationships are, qualitatively, as expected. The results are summarized in Table 1, and, once again, only modest gains are seen.

### 3.3.  DAOPHOT FIND

Sometimes, it is inconvenient to use the matched filter approach for object detection. DAOPHOT (Stetson 1987) trades away the detection efficiency of a matched filter for a simple convolution operation that helps to: 1) suppress galaxies and other extended objects, 2) recognize when seemingly extended objects are really a blend of two or more stellar images, and 3) compensate for a spatially varying background. This strategy differs from FOCAS, where objects (or groups of objects) are first detected with a matched filter and subsequently groups are split and classified.

Stetson's filter is a lowered, truncated Gaussian, possessing a FWHM equal to the final stellar FWHM and zero total volume.

$$W(x,y) = k \left[ G(x,y) - \sum_{i=1}^{m} G(x_i, y_i)/m \right] \qquad (10)$$

where:

$$G(x,y) = e^{-[(x-x_0)^2 + (y-y_0)^2]/2\sigma^2},$$

$\sigma$ is chosen such that $G(x,y)$ has the same FWHM as the PSF on the coadded frame, $m$ is the number of pixels one is summing over, and k is an arbitrary constant. The convolution sums are calculated centered on each data pixel in turn, extending to a radius of $1.5\sigma$ or 2 pixels whichever is larger.

In this case, because of the truncation, the calculations must be done numerically. However, they are analogous to the above calculations so we will omit the details.

## 3.4. Examples

The results of an example optimization for DAOFIND are shown in Fig. 2 and are summarized in Table 2. The s/n are substantially worse than the corresponding matched filter results. This is mainly due to the truncation, but there is also a contribution from the imperfect match between filter and signal. Extending the sum to larger radii would improve the s/n in an uncrowded field but would increase the possibility of contamination by neighboring objects in a crowded field, and make the convolution computationally more expensive.

For the case of variable seeing, there are two things worth noting: one is that the s/n enhancement due to the optimization (a $\sim 50\%$ increase is seen) is much greater than for the matched filter. The second is that the relationship between weighting and resolution, even for low-brightness objects, is non-monotonic, resulting in a final FWHM which is larger than the equal-weight FWHM. These effects both result from the pixel nature of the CCD coupled with the variable truncation radius of the convolving function. The truncation radius changes when we search through the weight parameter space, because the FWHM of the summed image is recalculated for each new set of guesses, and is in turn used to recalculate the truncation radius according to the above prescription. A minute increase in the FWHM, caused by adjusting the weighting, can result in four additional pixels being included in the sum and a discontinuous increase in the s/n. If CCDs provided a continuous image, the increased truncation radius would result in smooth changes in s/n, and a compromise would be established between signal increase and corresponding noise increase as a function of radius. Because of the discrete nature of the CCD, this compromise is always established at the point where the truncation radius just includes four additional pixels. This is because a compromise has been established between maximizing signal by including more pixels and increasing s/n by convolving with a narrower lowered Gaussian function. Therefore, the s/n enhancements are arising mainly from the inclusion of more pixels in the sum, and not from an improvement in overall image quality on the frame.

To explore this, we have done a similar calculation, but maintained the truncation radius appropriate to the FWHM of the equal-weight summed image. We have, however, allowed the $\sigma$ of the convolving Gaussian to vary such that its FWHM is equal to the FWHM of the co-added signal, as was done above. In this way, we sum over the same number of pixels as the equal-weight case, regardless of the current guesses, but, if the weighting results in reduced FWHM (which it does), we reduce the contamination and improve the star-galaxy discrimination. The results are shown in Fig. 3 and Table 2. The s/n improvements are similar to the variable truncation radius examples, but this time they are accompanied by decreases in the FWHM of up to 20%. Interestingly, the optimal weights now occupy a greater range of values; for the faintest example, the highest weight is more than 30 times larger than the lowest, compared to about a factor of four for the corresponding matched filter. The weighting will, however, be dependent on the number of pixels sampled by the convolution. Because of this result, and the fact that, in general, crowding and image resolution are uncorrelated, we feel it would be useful to add another user-adjustable parameter to DAOPHOT, namely a FIND radius parameter.

It is worth mentioning that we have not taken into account crowding effects on the s/n. These would increase the penalty against weightings which lead to larger final FWHM.

The results of the variable sky example, aside from lowered s/n, are very similar to the matched filter. Only modest s/n gains are seen despite a factor of three range in sky brightness.

We have not discussed the DAOFIND routines which conduct test for galaxies and stars since they are not relevant to our main goals. We have, however, included a discussion of alternative star/galaxy and star/cosmic ray discriminators in Appendix B..

## 4. Photometry

After detection, the usual next step is to carry out photometry, the process of measuring the relative brightnesses of previously detected objects. For stellar photometry this usually takes one of two forms: 1) aperture photometry, where all the flux within some radius of the star is summed, and/or 2) profile-fitting photometry, where the known stellar PSF is fit to the star with the scale (brightness) and position as the unknown parameters. The former is often sufficient in uncrowded regions while the latter is usually required in crowded fields.

### 4.1. Aperture Photometry

The optimal weighting for aperture photometry will be different than that for simple detection. The signal and noise expressions are given in eqns 1 and 2. We have neglected the small uncertainty in the local mean sky level. We wish to derive the optimal weights for adding images prior to photometry.

The integrated signal within a radius $R$ is:

$$S_I(R) = \int S(r)dA = \sum_{i=1}^{n} w_i I_i (1 - e^{-R^2/2\sigma_i^2}),\qquad(11)$$

where $r^2 = x^2 + y^2$, and the integrated variance is

$$[N_I(R)]^2 = \int [N(r)]^2 dA = \sum_{i=1}^{n} \left[ \frac{w_i^2 I_i}{g_i}(1 - e^{-R^2/2\sigma_i^2}) + \frac{(w_i^2 g_i B_i + w_i^2 h_i^2)\pi}{g_i^2} R^2 \right].\qquad(12)$$

The integrated s/n for aperture photometry is $S_I(R)/N_I(R)$. We have not taken into account the pixel nature of the CCD, however, this is not a serious problem for aperture photometry since we are usually dealing with the total flux in a large number of pixels and it is possible to include partial pixels. A pixel's contribution to the total flux is simply the pixel flux multiplied by the fraction of the pixel included within $R$.

For aperture photometry, we have another parameter, in addition to the weights $w_i$, to be optimized, namely the aperture radius $R$.

### 4.2.  Examples

The optimal weighting results for variable $\sigma_i$ are shown on the left side of Fig. 4. The results are summarized in Table 3, where column 1 contains the total stellar intensities, columns 2 and 3, the optimized s/n and FWHM, column 4, the optimal $R$, column 5, the ratio of $R$ over FWHM, and columns 6 and 7, the equal-weight s/n and FWHM. The right side of Fig 4. contains the results for variable sky which are summarized in Table 3. The gains in s/n and image resolution are similar to what was seen for the matched filter.

### 4.3.  Profile-Fitting Photometry

In a crowded stellar field (e.g. a globular cluster), it is generally advantageous to use profile-fitting photometry. The idea is to determine the PSF (or PSF as a function of position), and simultaneously fit it to stars which are sufficiently close to one another such that aperture photometry is unreliable. This is what is done in ALLSTAR (Stetson & Harris 1988), which utilizes a weighted linearized least-squares fitting routine to simultaneously determine the PSF scalings and centers for a group of stars. Linear least-squares, for a single object, is mathematically equivalent to convolution, so this process is similar to what we described in §3.1. However, in a weighted least-squares calculation, one weights each pixel according to a noise model which is dependent on the PSF scale, thereby increasing the s/n. Detailed discussions of least-squares fitting can be found in Stetson (1987) and references therein so we will restrict ourselves to the essentials.

We will consider only the case of an isolated star. Furthermore, we assume that the star center is perfectly determined (but see below) and lies at the center of a pixel. This represents the maximum s/n case. The signal on our co-added frame is:

$$S_A = \sum_{i=1}^{n} w_i I_i, \tag{13}$$

and the variance, or more correctly the standard error, squared is Press et al. (1986), p. 512,

$$\frac{1}{N_A^2} = \frac{1}{S_A} \sum_{r \le r_f} \frac{S_p(x, y)}{N_a(x, y)^2}, \tag{14}$$

where

$$N_a(x, y)^2 = N_p(x, y)^2 + \sum_{i=1}^{n} \frac{w_i^2}{g_i^2} \left\{ [0.0075 g_i (I_i + B_i)]^2 + \left( 0.027 \frac{g_i I_i}{\sigma_x^2 \sigma_y^2} \right)^2 \right\}.$$

$S_p(x, y)$ and $N_p(x, y)$ are given above. The two additional terms are from Stetson (1987); they are phenomenological error estimates for typical images. The first is an estimate of the flat-field error, and the second is the PSF interpolation error. Clearly, the s/n will increase as the area sampled increases. In practice, one must perform the fit over a small region specified by some *fitting radius*, $r_f$. The signal quoted here is an "aperture-corrected" signal while ALLSTAR returns only the signal within the fitting radius, however, the s/n are the same. Our noise calculation differs slightly from ALLSTAR in that the latter employs an additional weighting technique; the weighting is modulated by a radially decreasing function to prevent oscillating solutions. We have not included this term as it is not an uncertainty, but a byproduct of the pixelisation.

## 4.4.  Example

Fig. 5 shows the results of the tests done for the profile-fitting photometry which are summarized in Table 4. For these tests, the fitting radius was set to the FWHM of the coadded image. The s/n improvements we see in our example are quite small. The image resolution improves by about 10% in the best case, this is increased if one holds the fitting radius constant. Interestingly, the highest s/n example does not exhibit a monotonic weight vs. $\sigma$ relationship, similar to the matched filter. This is not simply a consequence of a variable fitting radius coupled with pixelisation as was the case for DAOFIND, since this effect is also present when the fitting radius is fixed at a constant value.

In practice, centroiding errors will decrease the s/n. We do not feel it is necessary to include these effects for two reasons. The first is that centroiding errors will only have a second order effect on the fitted value of the flux. An off-center PSF will underestimate the flux in the direction of the true center while overestimating the flux in the opposite direction resulting in only a small net bias. The second is that since the optimal weighting decreases the FWHM on the final frame, the centering errors should be smaller which should also help to improve the s/n resulting in a higher percentage increase in the s/n.

## 5.  The Program

We have written a Fortran program employing the downhill Simplex method to calculate the optimal weighting for the four scenarios outlined above. It produces weights that can be applied directly to the individual CCD frames. This program is available by emailing philf@physics.att.com. Include in your letter an internet FTP site with username and password. The simplest option would be to specify an anonymous FTP site. The program comes with complete documentation.

## 6.  Conclusion

In this paper we have described processes for the optimal co-addition of CCD frames containing stellar fields. Optimal has been defined to mean the maximizing of signal-to-noise for four different goals. These are 1) convolution with a matched filter of infinite extent, 2) convolution with a truncated, lowered Gaussian of zero volume (DAOFIND), 3) aperture photometry, and 4) profile-fitting photometry (ALLSTAR). We carried out several examples involving 21 individual frames for which either the sky brightness or the image resolution varied by a factor of three.

1) The gains in s/n, over the equal-weight case, were modest for the specific examples described in the text, except for variable $\sigma$ DAOFIND which exhibited substantial gains.

2) The image resolution on the weighted frame improved over the equal-weight case for most of the variable $\sigma$ examples, except for the highest signal matched filter and ALLSTAR, and the variable sampling radius DAOFIND.

3) In Fig. 6 we show a superposition of the asymptotic weightings for the four different measurement techniques described above. Aside from DAOFIND (see §3.3.) the weightings agree well for variable $\sigma$. The agreement is perfect for all four techniques with variable sky (uniform $\sigma$) and has the analytic solution $w_i \propto g_i^2 I_i (g_i B_i + h_i^2)^{-1}$, as $I_i$ tends to zero ($w_i$ is the weight of frames $i$, $I_i$ is the total stellar flux, $g_i$ is the gain, $B_i$ is the sky brightness, and $h_i$ is the readout noise.

4) We have made empirical fits in the small $I_i$ limit for the function $w_i = a(1 + \sigma_i)^b g_i^2 I_i r^2 (g_i B_i + h_i^2)^{-1}$, for $1 \leq \sigma_i$ (pix) $\leq 3$ and $g_i B_i + h_i^2$ varying by a factor of three for the aperture photometry optimization (should be valid for all but DAOFIND). This phenomenological function is simply the above analytical solution times a reasonable guess at the $\sigma$ dependence. We found that $b \approx -2.0$. In this range, the residuals were less than 10% of the $w_i$.

5) The gain in s/n, relative to the equal weight case, is roughly proportional to the variance of $\log(w_i)$.

6) Real CCD images of stars do not have perfectly Gaussian profiles, possessing broader wings and sharper cores. The question arises, how does this change the optimization. We have performed a test with a PSF consisting of two Gaussians, the second possessing a width 2.45 times larger than the first, and a peak, a tenth that of the first. This is a typical value seen in images (Tonry 1987). Using the aperture photometry algorithm with variable $\sigma$ and total stellar intensities equal to 200 DU (as was done with the Gaussian PSF) we found that the optimal weights differed by much less than 1% from the Gaussian PSF optimal weights. The s/n improved from 4.2 to 4.5 and the FWHM decreased from 3.69 to 3.35, very similar percentage changes to the Gaussian PSF case (6.5 to 7.0 and 3.47 to 3.16). We conclude that more realistic PSFs, while having systematically lower s/n, will have similar optimal weighting.

Finally, a word about galaxies is in order. The appearance of resolved, or partially resolved, faint galaxies is less sensitive to seeing conditions than is the appearance of a star. As we have seen, the s/n ratios for stars are not strongly dependent on the weighting scheme, and we expect galaxy s/n to be even less dependent. Therefore, it is probably valid to ignore the $\sigma_i$ for most photometry of faint galaxies, setting them all to some characteristic value which produces signals similar to the galaxies of interest. Alternatively, it would be possible to modify WEIGHT to use galaxy profiles such as $R^{1/4}$ profiles, which could then be convolved with a blurring function to represent the seeing on each frame.

We would like to acknowledge Peter Stetson for some very useful email correspondences and for his comments on an earlier version of the manuscript. We would also like to thank Tony Tyson for many useful comments and suggestions PF thanks NSERC and Bell Laboratories for postdoctoral fellowships. Partial support for this work was provided by NASA through grant number GO-2684.09-87A from the Space Telescope Science Institute, which is operated by the Association of Universities for Research in Astronomy, Inc., under NASA contract NAS5-26555.

## A.  Cosmic Ray and Defect Rejection

If one wants to employ clipping algorithms to reject cosmic rays, the weighting schemes outlined in this paper may need to be modified. This is most likely to occur if there are only a few frames or the number of cosmic rays is very high. In cases of few frames, a very small subset (or even one frame) may posses over half the weight. If cosmic rays were to fall in the same place on these highly weighted frames they would be incorporated in the final image. Therefore, there may be situations where wants to reduce the weights of the most highly weighted frames, sacrificing some s/n gains for improved cosmic ray rejection.

We consider a generalized weighted median, where one clips a fraction $f$ of the weight of the data from each tail, and then averages the remaining central portion. This includes both weighted average (f=0), and a weighted median (f=0.5) as limiting cases. The sky backgrounds must have been equalized, and the frames normalized prior to employment of the clipping algorithm.

Assume we have $N$ pixels and $R$ cosmic rays per each of $n$ images. The probability of finding cosmic rays on the same pixel in $K$ separate images is

$$P(K) = \frac{n!}{K!(n-K)!} \left(\frac{R}{N}\right)^K ,$$  (A1)

In the equal weight case, we want to set our clipping fraction such that it can clip $K_{max}$ pixels where $P(K_{max}) = F/N$, thus guaranteeing that on average there will be only $F$ pixels on the final image contaminated by cosmic rays. $K_{max}$ is easily found by using either Stirling's approximation or an iterative fixed point method, and $f$ must be set greater than $K_{max}/n$. For example, with $N = 10^6, n = 10, R = 100, F = 1.0$ one obtains $K_{max} = 2.0$, $f = K_{max}/N = 0.2$; clipping two pixels from each tail and summing the remaining six pixels.

The weighted case is slightly more complicated. The most conservative approach is to assume that the cosmic rays are falling on the pixels with the highest weight, then:

$$f \geq \sum_{i}^{K_{max}} w_i, \tag{A2}$$

where the $w_i$ are summed in order of their value, from highest to lowest. Since there is an upper limit to $f$ of one half, our technique of optimized weights may make it difficult, in some cases, to maintain adequate cosmic ray rejection. If you are willing to sacrifice s/n for improved cosmic ray rejection then the solution is to reduce the range of the weights until $f \leq 0.5$.

## B. Optimal Star/Galaxy or Star/Cosmic Ray Discrimination

Using techniques in Whalen 1971, one can calculate the optimal filter for discriminating between stars and galaxies, or stars and cosmic rays, assuming an uncrowded field and faint objects. The optimal filter for discriminating between objects with PSFs $A$ and $B$ is $D = \alpha A - B$, where $\alpha$ is chosen so that $D \star A > 0$ and $D \star B < 0$, and 0 is used as the decision point. If one wants it to be equally probable that both galaxies and stars will be misclassified, one sets $D \star A = -D \star B$, and therefore, $\alpha = (B \star B + A \star B)/(A \star A + A \star B)$.

The optimal discriminator between a star and a barely resolved galaxy can be calculated by taking $A$ to be a Gaussian PSF of size $\sigma$, and $B$ to be that Gaussian convolved with a galaxy of size $r_s$. The continuous approximation to the star-galaxy discrimination filter in the limiting case of small galaxies is

$$D_{sg} = \lim_{r_s \to 0} \frac{D}{r_s^2} \sim (\sigma^2 - r^2)e^{-r^2/2\sigma^2}. \tag{B3}$$

Qualitatively, this is similar to the DAOFIND detection filter: a Gaussian-like center with negative wings, however the DAOFIND filter does not discriminate very strongly against galaxies; it gives a positive result when convolved with any size galaxy. $D_{sg}$ gives a negative result when convolved with anything larger than a star. Increasing our proposed FIND radius for DAOFIND yields functions that are reminiscent of the optimal filters for discriminating stars from larger galaxies: sombrero-like functions with a high Gaussian center, and broad, low, negative tails.

If a cosmic ray is taken to be a single isolated hot pixel, one can similarly calculate the optimal filter to discriminate between stars and cosmic rays. Not surprisingly, it is the difference between a delta-function and the stellar PSF. As there is no elegant analytic form for $\alpha$ in this case, we recommend that it be calculated numerically. This discriminator is qualitatively similar to the DAOFIND "sharp" statistic, but it will have a higher s/n.

| $I_i$ | Optimal Weights | | Equal Weights | |
|---|---|---|---|---|
| | s/n | FWHM | s/n | FWHM |
| (DU) | | (pix) | | (pix) |
| | Variable $\sigma$ | | | |
| 200.0 | 7.7 | 3.18 | 7.3 | 3.47 |
| 2000.0 | 74.3 | 3.21 | 71.2 | 3.47 |
| 5000.0 | 175.5 | 3.26 | 170.2 | 3.47 |
| 20000.0 | 573.8 | 3.40 | 570.4 | 3.47 |
| | Variable Sky | | | |
| 200.0 | 9.5 | 2.35 | 10.0 | 2.35 |
| 2000.0 | 95.3 | 2.35 | 91.4 | 2.35 |
| 5000.0 | 221.3 | 2.35 | 214.4 | 2.35 |
| 20000.0 | 685.9 | 2.35 | 678.0 | 2.35 |

Table 1: Matched Filter

| $I_i$ | Optimal Weights | | Equal Weights | |
|---|---|---|---|---|
| | s/n | FWHM | s/n | FWHM |
| (DU) | | (pix) | | (pix) |
| Variable $\sigma$ – Variable Truncation Radius | | | | |
| 200.0 | 2.4 | 3.51 | 1.6 | 3.47 |
| 2000.0 | 23.2 | 3.51 | 16.0 | 3.47 |
| 5000.0 | 55.1 | 3.51 | 38.1 | 3.47 |
| 20000.0 | 181.0 | 3.51 | 126.7 | 3.47 |
| Variable $\sigma$ – Constant Truncation Radius | | | | |
| 200.0 | 2.4 | 2.82 | 1.6 | 3.47 |
| 2000.0 | 22.5 | 2.84 | 16.0 | 3.47 |
| 5000.0 | 52.1 | 2.86 | 38.1 | 3.47 |
| 20000.0 | 159.2 | 2.92 | 126.7 | 3.47 |
| Variable Sky | | | | |
| 200.0 | 4.8 | 2.35 | 4.6 | 2.35 |
| 2000.0 | 46.0 | 2.35 | 44.1 | 2.35 |
| 5000.0 | 104.2 | 2.35 | 107.7 | 2.35 |
| 20000.0 | 340.9 | 2.35 | 336.3 | 2.35 |

Table 2: DAOFIND

| $I_i$ | s/n | Optimal Weights | | | Equal Weights | |
| | | FWHM | $R$ | $R$/FWHM | s/n | FWHM |
| (DU) | | (pix) | | (pix) | (pix) | |
| | | | Variable $\sigma$ | | | |
| 200.0 | 7.0 | 3.16 | 2.3 | 0.72 | 6.5 | 3.47 |
| 2000.0 | 67.2 | 3.18 | 2.4 | 0.75 | 63.2 | 3.47 |
| 5000.0 | 159.9 | 3.22 | 2.5 | 0.78 | 152.6 | 3.47 |
| 20000.0 | 540.6 | 3.32 | 3.0 | 0.90 | 530.8 | 3.47 |
| | | | Variable Sky | | | |
| 200.0 | 9.5 | 2.35 | 1.6 | 0.68 | 9.0 | 2.35 |
| 2000.0 | 90.2 | 2.35 | 1.6 | 0.70 | 86.4 | 2.35 |
| 5000.0 | 211.3 | 2.35 | 1.7 | 0.72 | 204.2 | 2.35 |
| 20000.0 | 678.1 | 2.35 | 1.9 | 0.81 | 667.5 | 2.35 |

Table 3: Aperture Photometry

| $I_i$ | Optimal Weights | | Equal Weights | |
|---|---|---|---|---|
| | s/n | FWHM | s/n | FWHM |
| (DU) | | (pix) | | (pix) |
| | Variable $\sigma$ | | | |
| 200.0 | 7.6 | 3.18 | 7.2 | 3.47 |
| 2000.0 | 73.1 | 3.21 | 69.6 | 3.47 |
| 5000.0 | 173.1 | 3.24 | 166.5 | 3.47 |
| 20000.0 | 565.5 | 3.34 | 557.3 | 3.47 |
| | Variable Sky | | | |
| 200.0 | 8.5 | 2.35 | 7.8 | 2.35 |
| 2000.0 | 84.8 | 2.35 | 79.1 | 2.35 |
| 5000.0 | 208.6 | 2.35 | 199.0 | 2.35 |
| 20000.0 | 643.4 | 2.35 | 630.0 | 2.35 |

Table 4: Profile-Fitting Photometry

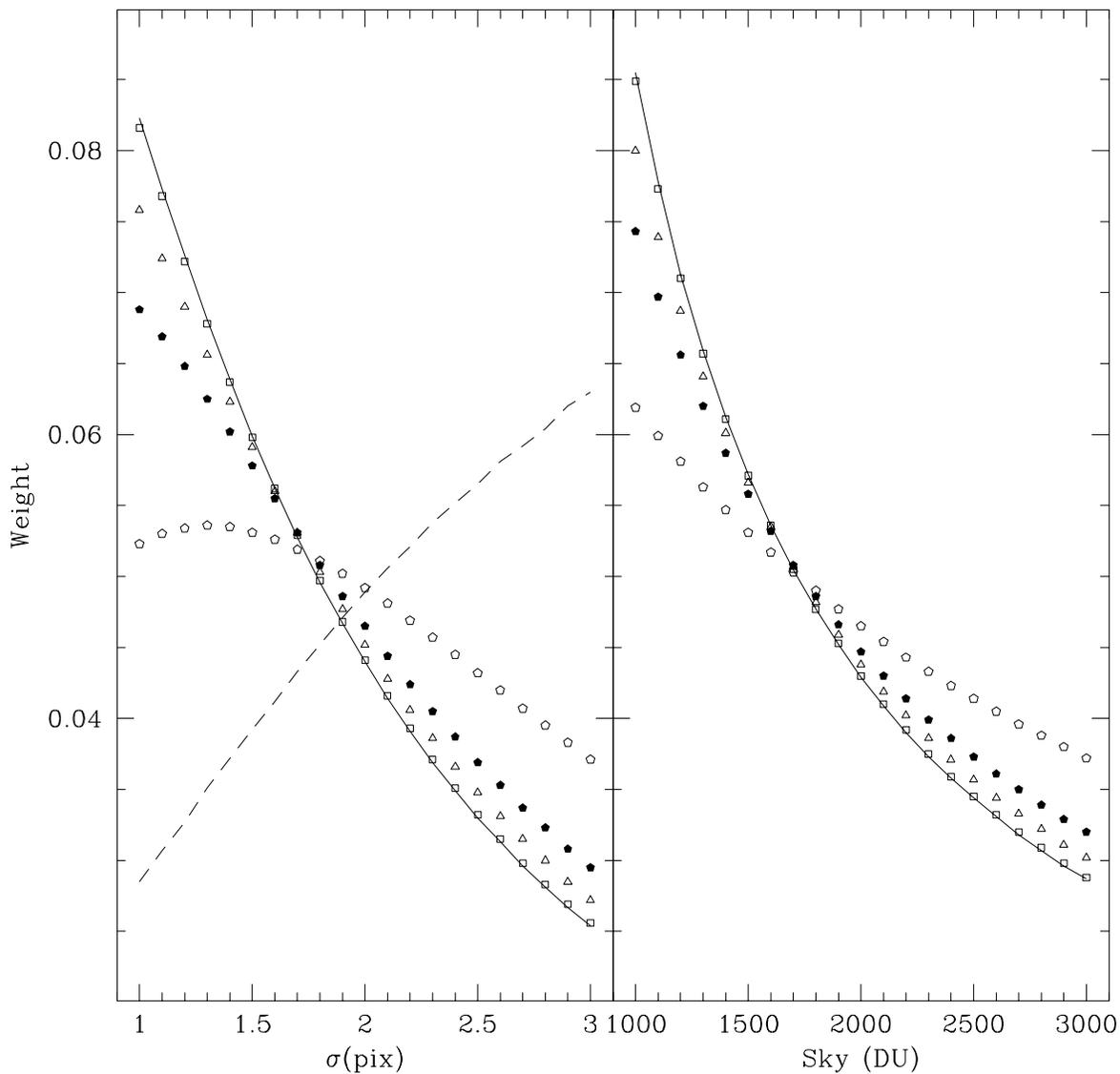

Fig. 1.— Weighting for the matched filter. On the left is a plot of weighting vs. $\sigma$. The squares are for $I_i = 200$ DU, triangles are $I_i = 2000$ DU, filled pentagons are $I_i = 5000$ DU and open pentagons are $I_i = 20000$ DU. The sky was 1000 DU for all cases. The solid line represents the weighting as the signal approaches zero and the dashed line is the relationship for zero sky and readout noise. The right side is a plot of weighting vs. sky brightness. The points are analogous to the left side. Values for the other parameters can be found in the text.

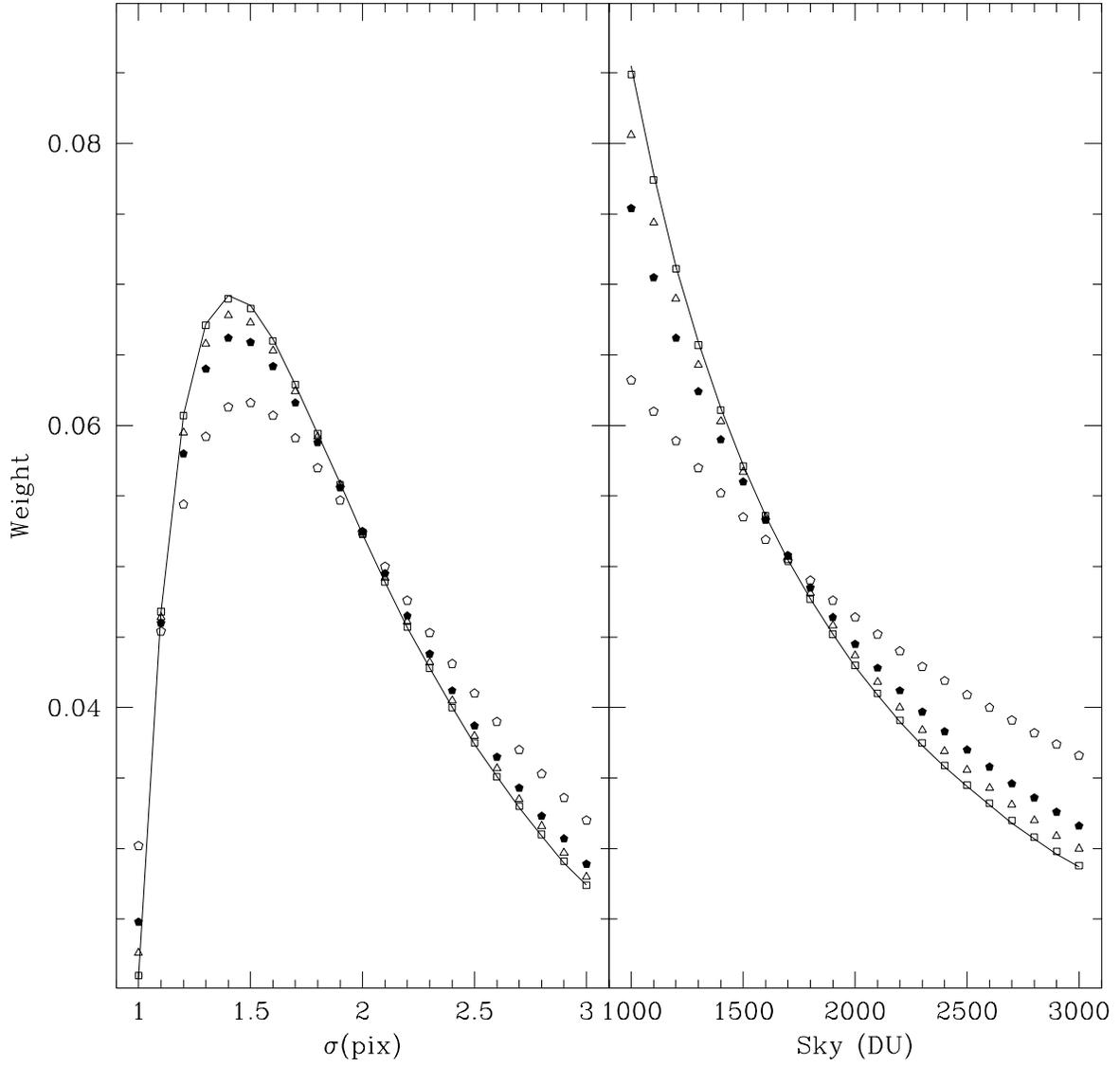

Fig. 2.— Weighting for the DAOFIND. See Fig. 1 for description.

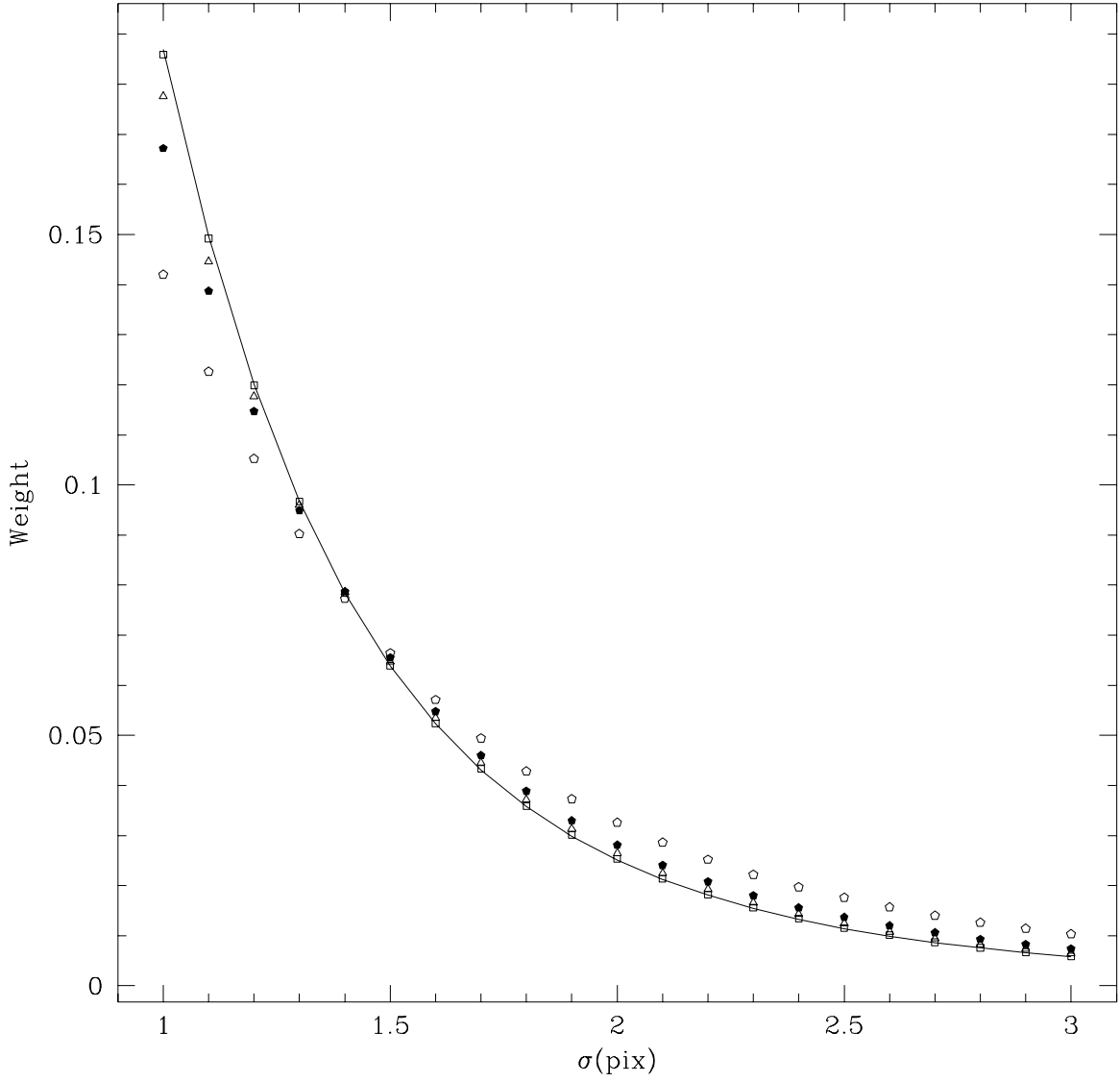

Fig. 3.— Weighting for the DAOFIND - Constant sampling radius. See Fig. 1 for description.

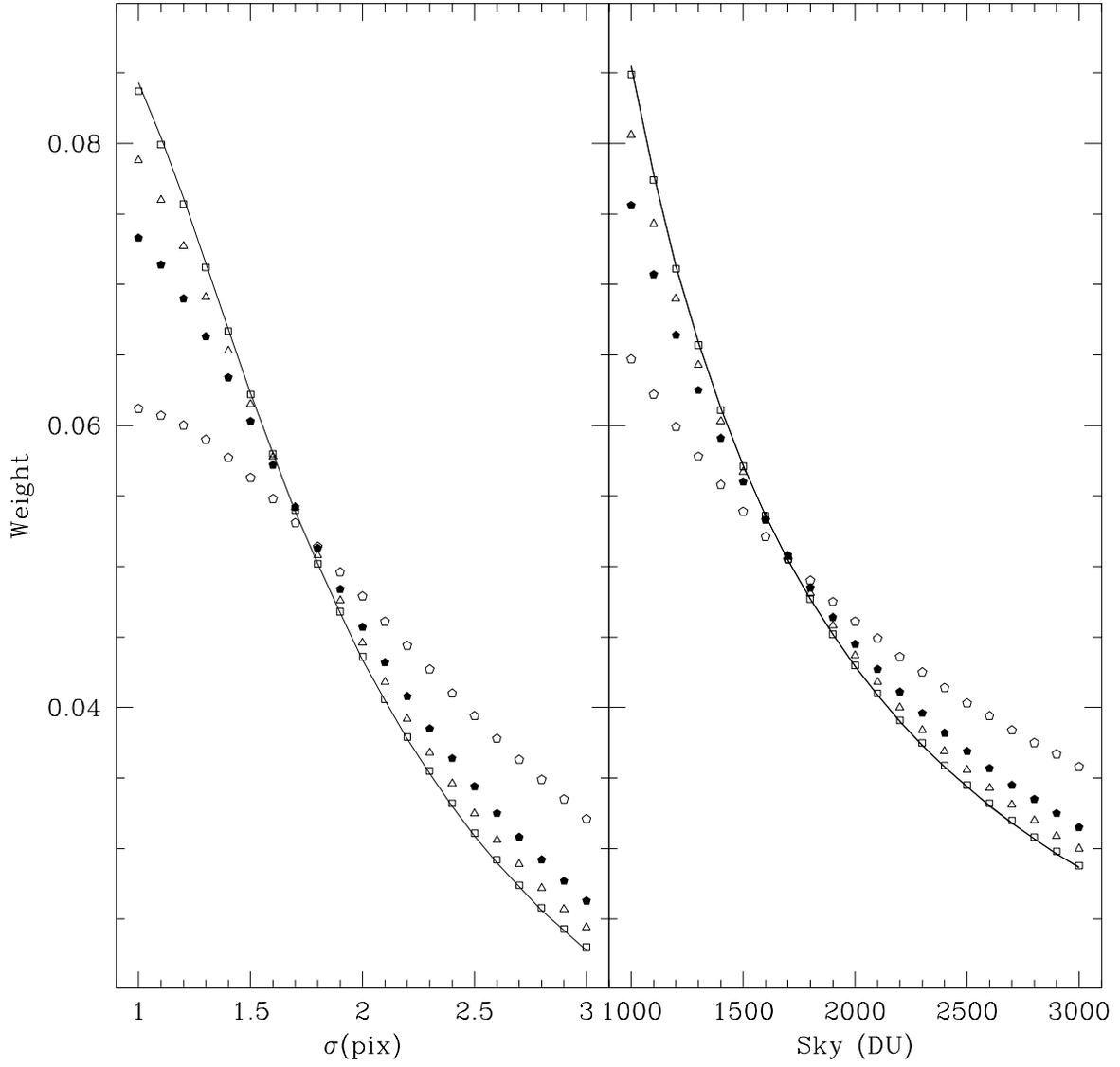

Fig. 4.— Weighting for aperture photometry. See Fig. 1 for description.

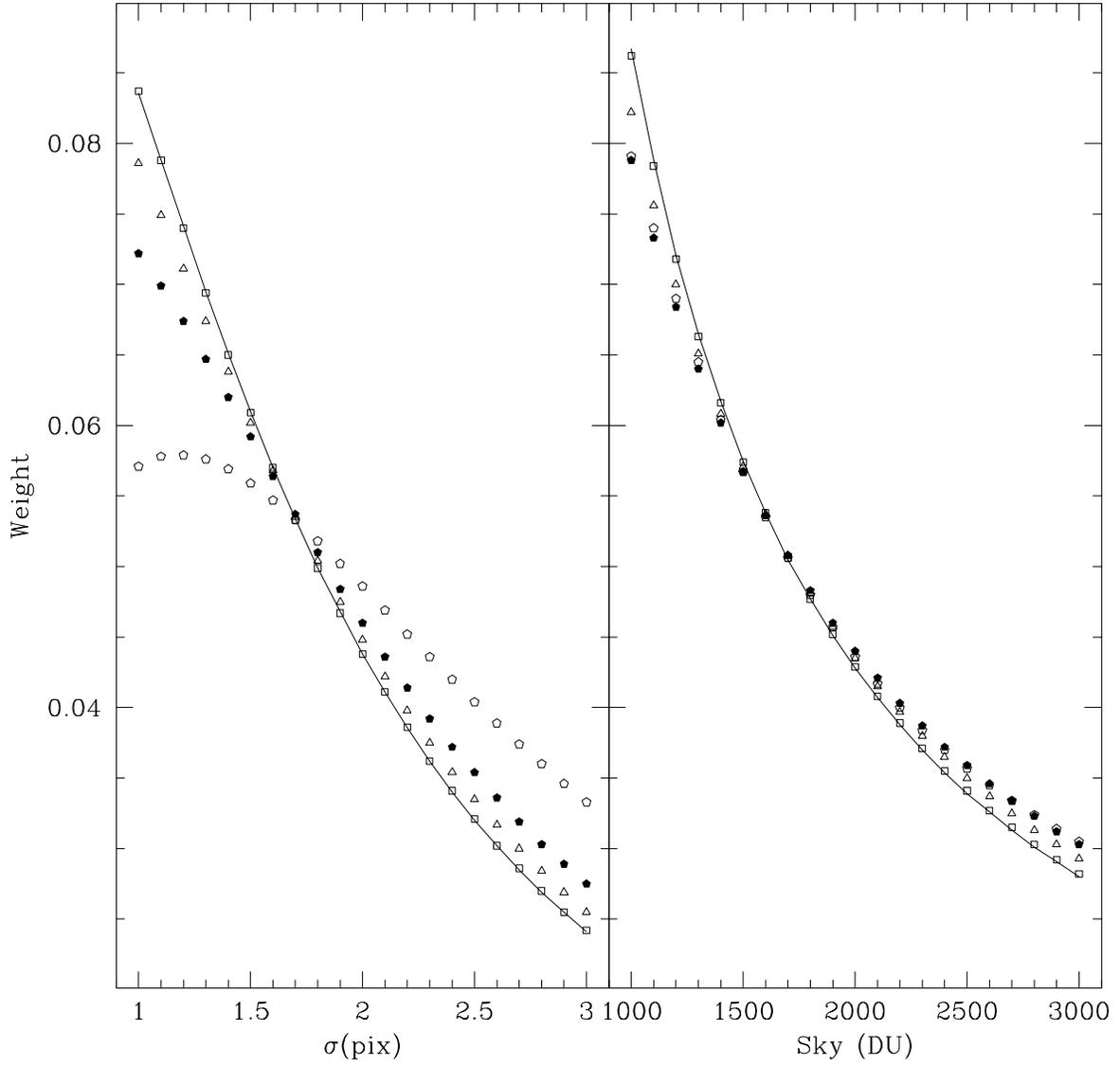

Fig. 5.— Weighting for ALLSTAR. See Fig. 1 for description.

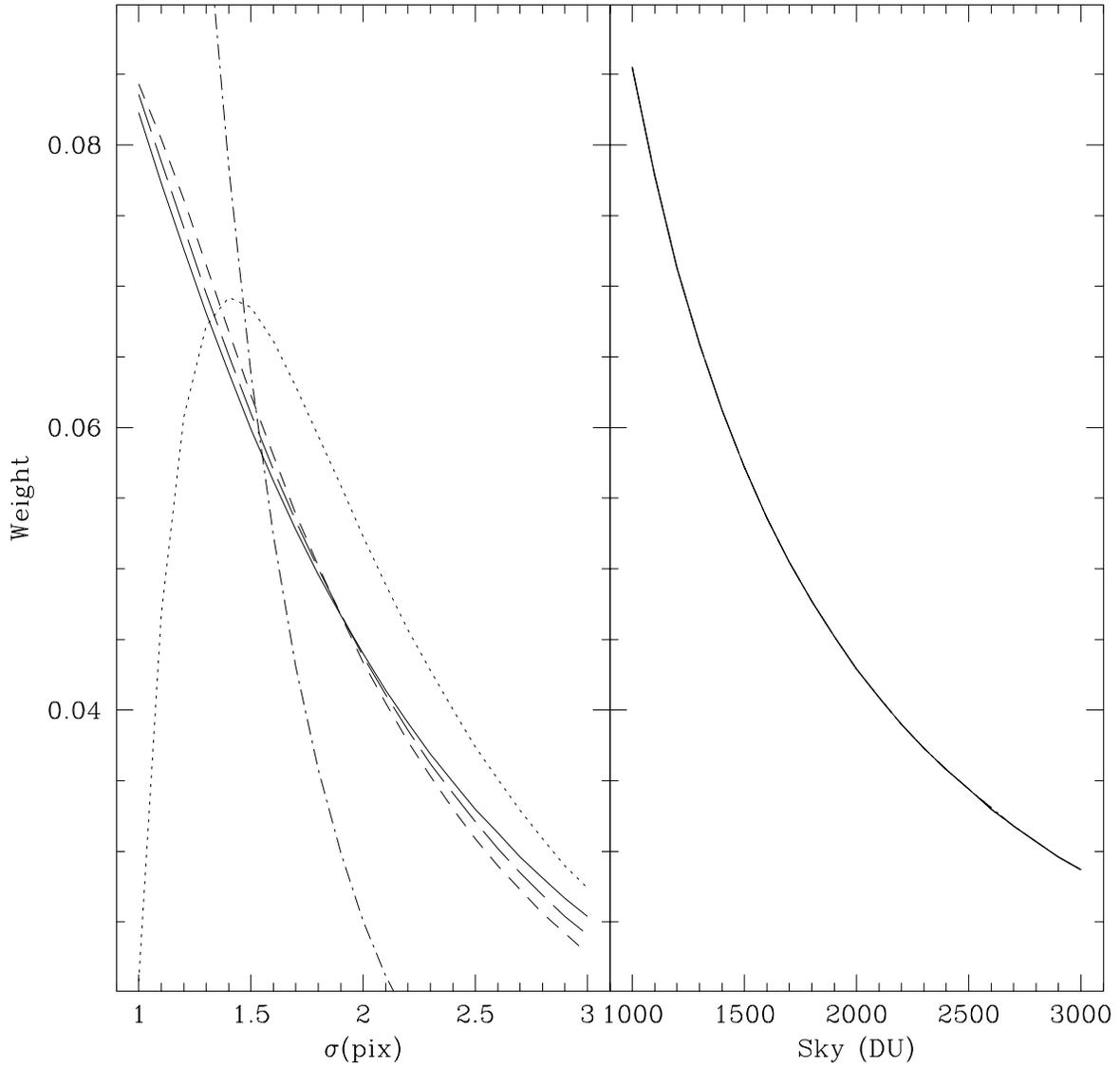

Fig. 6.— Comparison of weighting at the Low brightness limit. The left side shows the variable $\sigma$ result, and the right side is the variable sky result. The solid line is the matched filter, the dotted line is DAOFIND, the dot-short dash line is DAOFIND with constant sampling radius, the short dash is aperture photometry, the long dash is ALLSTAR photometry. The lines are virtually coincident for the variable sky case.